# Enhanced electron transfer using NiCo$_2$O$_4$@C hollow nanocages with an electron-shuttle effect for efficient tetracycline degradation

Yuwen Chen[a], Ke Zhu[a], Wenlei Qin[a], Zhiwei Jiang[a], Zhuofeng Hu[a], Mika Sillanpää[b], Kai Yan[a], *

[a] *Guangdong Provincial Key Laboratory of Environmental Pollution Control and Remediation Technology, School of Environmental Science and Engineering, Sun Yat-sen University, Guangzhou 510275, China.*

[b] *Department of Biological and Chemical Engineering, Aarhus University, Nørrebrogade 44, 8000 Aarhus C, Denmark.*

*Corresponding author. Email address: yank9@mail.sysu.edu.cn (K. Yan)

**Abstract:** Spinel oxides are recognized as promising Fenton-like catalysts for the degradation of antibiotics. However, the catalytic performance is restrained by the poor electron transfer rate (ETR). Herein, hollow NiCo$_2$O$_4$@C nanocages are rationally designed and prepared to accelerate ETR in peroxymonosulfate (PMS) activation for tetracycline (TC) degradation. Enhanced ETR of the NiCo$_2$O$_4$@C/PMS system is due to three aspects: (1) The hollow nanocage facilitates the diffusion and adsorption of TC, improving the ion transfer at a macroscopic level; (2) Electron reconfiguration in octahedral sites of NiCo$_2$O$_4$ increases the ratio of Co$^{2+}$, resulting in highly efficient PMS activation; (3) In-situ generated carbon acts as "electron shuttles", improving the electrical conductivity of catalysts at a microscopic level. As a result, the NiCo$_2$O$_4$@C demonstrates rapid ETR, leading to a high-efficiency activation of PMS. The NiCo$_2$O$_4$@C/PMS system exhibits exceptional TC degradation efficiency and reusability. Non-radical pathway, including $^1$O$_2$ and direct electron transfer, dominates the system. This work offers a feasible strategy for enhancing electron transfer in the Fenton-like system.

**Keywords**: Spinel oxide; Electron shuttle; Electron transfer; Tetracycline; Peroxymonosulfate

## 1. Introduction

The excessive usage of antibiotics including tetracycline (TC) [1] and improper wastewater treatment lead to emerging pollution and threat to human health [2,3]. Thus, it is urgent to develop innovative approaches to tackle this issue. Heterogeneous Fenton-like catalysis based on peroxymonosulfate (PMS) is considered one of the most effective techniques [4,5] with high oxidation capacity and minimized catalyst loss for the removal of organic pollutants. However, the slow electron transfer rate (ETR) limits the efficiency of PMS activation and antibiotic removal [6,7]. Therefore, the key to improving the efficiency of a Fenton-like system relies on the higher ETR.

At a macroscopic level, ETR can be improved by improving the ion transfer process, including diffusion and adsorption of PMS and antibiotics. Various strategies have been exploited. For example, Liu et al. [8] assembled $Co_3O_4$ in a carbon nanotube to create a nanoconfined structure, which can enrich contaminants and mitigate the mass transfer resistance. At a microscopic level, the electrical conductivity of catalysts is crucial for improving ETR. For metal-based catalysts, especially, higher electronic conductivity means a faster redox cycle among multivalent metals [9], resulting in higher PMS activation efficiency. Based on these thoughts, novel Fenton-like catalysts could be rationally designed.

Spinel oxides, which consist of metal-oxygen octahedron and tetrahedron structures, have been widely applied in pollutant degradation [10,11]. Compared to monometallic spinel oxides, bimetallic spinel oxides have higher intrinsic activity and stability owing to the synergy of different metals. Moreover, electron reconfiguration between different metals in spinel oxides has been widely observed in electrocatalysis [12,13]. Li et al. reported [14] that Ni dopants facilitate the formation of oxygen vacancies and enhance electron transfer from $Ni^{2+}$ to $Co^{3+}$. It is reasonable to assume that such electron interactions may also promote PMS activation. However, the poor electrical conductivity of spinel oxides hinders the ETR, leading to insufficient PMS activation and instability of catalysts. In this case, carbonaceous materials such as biomass carbon, graphite, and carbon nitride are good candidates for improving ETR and stabilizing spinel oxides [15–17]. Carbonaceous materials typically exhibit high electrical conductivity due to the abundance of free electrons that can move among delocalized π bonds formed by $sp^2$-hybridized carbon atoms. These free electrons in carbon can make it serve as "electron shuttles", mediating electron transfer in the Fenton-like system [18]. Therefore, combining spinel oxides with carbon is a rational strategy for efficient Fenton-like degradation of antibiotics.

Conventional modification needs extra carbon sources as support, which may exhibit a weaker anchoring effect for spinel oxides. Metal-organic frameworks (MOFs) are considered ideal precursors for preparing metal oxides [19–21]. The metals coordinated to organic ligands are arranged uniformly, preventing the agglomeration of metal oxides during calcination. Besides, organic ligands can serve as carbon sources, generating carbon in situ. These in-situ generated carbon has a more robust interaction with metal oxides than external carbon support which leads to higher ETR. Moreover, MOFs usually possess a unique structure that can be exploited to create a nanoconfined structure for enrichment of antibiotics.

In this work, we utilized zeolitic imidazolate framework-67 (ZIF-67) as a template to prepare NiCo$_2$O$_4$@C hollow nanocages with in-situ generated carbon to enhance the ETR of the Fenton-like system. The rapid ETR of NiCo$_2$O$_4$@C was attributed to the electron reconfiguration and electron-shuttle effect. Initially, we examined the unique nanocage structure and the composition of the catalysts to confirm the successful preparation of NiCo$_2$O$_4$@C. The chemical state of NiCo$_2$O$_4$@C was further elucidated. Next, the catalytic performance of NiCo$_2$O$_4$@C hollow nanocages was assessed by activating PMS to degrade TC. Later, the contribution of radical and non-radical pathways was examined using various techniques, including chemical probe methods and electrochemical measurements. With the assistance of theoretical calculations, the toxicity of degradation intermediates was reasonably assessed.

## 2. Experimental section

*2.1. Chemicals*

Co(NO$_3$)$_2$·6H$_2$O, tetracycline (TC, C$_{22}$H$_{24}$N$_2$O$_8$) and furfural alcohol (FFA, C$_5$H$_6$O$_2$) were purchased from Aladdin. Ni(NO$_3$)$_2$·6H$_2$O, 2-methylimidazole (2-MeIm, C$_4$H$_6$N$_2$), oxytetracycline (OTC, C$_{22}$H$_{24}$O$_9$N$_2$), chlortetracycline hydrochloride (CTC, C$_{22}$H$_{23}$ClN$_2$O$_8$·HCl), benzoic acid (BA, C$_6$H$_5$COOH) and nitrobenzene (NB, C$_6$H$_5$NO$_2$) were purchased from Macklin. Peroxymonosulfate (PMS, KHSO$_5$·0.5KHSO$_4$·0.5K$_2$SO$_4$) was purchased from Sigma Aldrich.

*2.2. Catalyst preparation*

In a typical synthesis of ZIF-67 [22], 2.9 g Co(NO$_3$)$_2$·6H$_2$O and 3.3 g 2-MeIm were separately dissolved in 100 mL of methanol. The former was added dropwise into the latter under stirring, and the final suspension was kept for 24 h. The purple suspension was then rinsed with methanol, centrifuged, and dried in a vacuum at 60 °C for 12 h.

ZIF-derived NiCo$_2$O$_4$@C hollow nanocages were prepared through ion exchange and calcination procedures (Fig. 1). In detail, as-prepared ZIF-67 and Ni(NO$_3$)$_2$·6H$_2$O at a mass ratio of 1:2 were mixed in 100 mL of absolute ethanol and stirred vigorously for 30 min, in which Ni$^{2+}$ was partially substituted for Co$^{2+}$ in the ZIF-67 framework by ion exchange. The etched ZIF-67, denoted as NiCo-ZIF, was afterward collected by filtration. It was rinsed with ethanol and dried in a vacuum. Later, the powder was placed in a combustion boat and carbonized at 600 °C under an Ar atmosphere with a heating rate of 1 °C min$^{-1}$ for 2 h. The carbonized materials were then calcined again in the air at 350 °C with the same ramp rate for 2 h to obtain NiCo$_2$O$_4$@C. In addition, ZIF-derived Co$_3$O$_4$ and NiCo$_2$O$_4$ were prepared by directly calcinating ZIF-67 and NiCo-ZIF in the air at 350 °C, respectively.

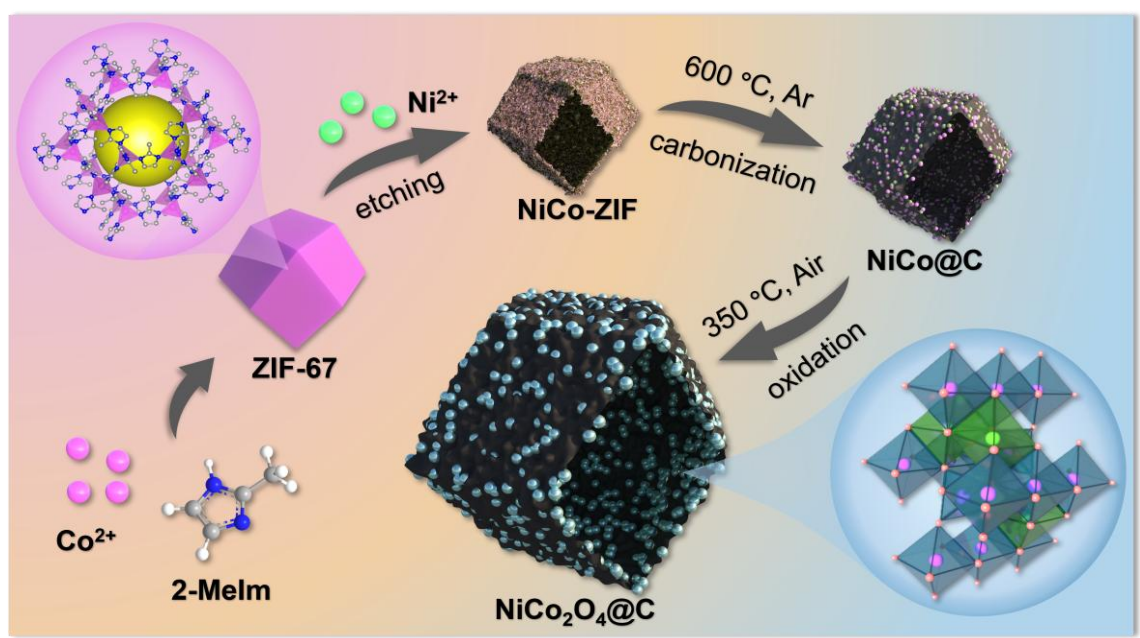

**Fig. 1**. Synthesis routine of ZIF-derived Co$_3$O$_4$, NiCo$_2$O$_4$ and NiCo$_2$O$_4$@C.

*2.3. Characterization*

X-ray diffraction (XRD, Rigaku Ultima IV) was used to investigate the composite of catalysts. Scanning electron microscopy (SEM, ZEISS Sigma 500) and transmission electron microscope (TEM, FEI Talos-F200S) were used to observe the morphologies.

An automatic surface area and porosity analyzer (BET, Micromeritics TriStar 3000) were employed to analyze the specific surface area. Raman spectrometer (Renishaw inVia) was employed with 532 nm HeNe lasers. Fourier transform infrared spectroscopy (FT-IR, Nicolet iS50) was employed. Electron paramagnetic resonance (EPR) was performed on Bruker A300-10-12. The surface element valence state of catalysts was examined with K-Alpha X-ray Photoelectron Spectrometer (XPS). Electrical conductivity was measured by a four-probe resistance ratio tester (FT-8100A). The leaching level of metals was examined by inductively coupled plasma-optical emission spectrometry (ICP-OES, PerkinElmer Optima 5300DV).

*2.4. Electrochemical measurements*

Electrochemical methods were performed on a CHI761E electrochemical workstation. The working electrode was prepared in this way: 5 mg of catalysts were dispersed in 5% Nafion ethanol solution under sonication for 10 min. 100 μL of the suspension was added dropwise on an FTO glass and dried at room temperature. A conventional three-electrode system was adopted. Pt wire was used as the counter electrode and Ag/AgCl was employed as the reference electrode. The electrolyte was 1 M $Na_2SO_4$. Linear sweep voltammetry (LSV) was assessed at 5 mV $s^{-1}$. Chronoamperometry measurements were carried out under 0.4 V vs Ag/AgCl.

*2.5. Experiment procedures*

TC degradation experiments were performed at room temperature in 50 mL of 10 mg $L^{-1}$ TC solution. 5 mg catalysts were dispersed in the solution with ultrasonication followed by magnetic stirring for 60 min to achieve adsorption-desorption equilibrium, then 0.8 mM PMS was injected to initiate the degradation reaction. 0.5 mL of samples were withdrawn at specific intervals within 20 min, which were immediately mixed with 0.5 mL of methanol to terminate the reaction.

The concentration of collected samples were determined by high-performance liquid chromatography (HPLC, Shimadzu LC-20A). A Shimadzu-C18 column (4.6 × 250 mm, 5 μm) was deployed to elute target pollutants. Detection conditions were listed in Table S1. Degradation intermediates were analyzed by liquid chromatography-mass spectroscopy (LC-MS, Thermos Scientific Ultimate 3000 HPLC). The mobile phase comprised 0.1% formic acid water solution and 0.1% acetic acid acetonitrile solution at 0.3 mL min$^{-1}$ under gradient elution mode.

*2.6. Steady-state concentration of reactive oxygen species*

The steady-state concentrations of reactive oxygen species (ROS) were measured and calculated with the chemical probe method reported in [23]. Specifically, BA, NB and FFA were used as chemical probes (0.1 mM for BA, 0.2 mM for NB and FFA) to capture different ROS. The change in probe concentrations was detected with HPLC (Table S1) and can be expressed in Eqs. 1-3:

$$-ln\frac{[BA]}{[BA]_0} = \left(k_{BA,\cdot OH}[\cdot OH]_{ss} + k_{BA,\cdot SO_4^-}[\cdot SO_4^-]_{ss}\right)t = k_{obs,BA}\, t \quad (1)$$

$$-ln\frac{[NB]}{[NB]_0} = \left(k_{NB,\cdot OH}[\cdot OH]_{ss} + k_{NB,\cdot SO_4^-}[\cdot SO_4^-]_{ss}\right)t = k_{obs,NB}\, t \quad (2)$$

$$-ln\frac{[FFA]}{[FFA]_0} = \left(k_{FFA,\cdot OH}[\cdot OH]_{ss} + k_{FFA,\cdot SO_4^-}[\cdot SO_4^-]_{ss} + k_{FFA,^1O_2}[^1O_2]_{ss}\right)t$$
$$= k_{obs,NB}\, t \quad (3)$$

In the equations, $k_{probe,ROS}$ represents the second-order reaction rate constants between probes and ROS (M$^{-1}$ s$^{-1}$, Table S2), while $k_{obs,probe}$ represent the observed pseudo-first-order reaction rate constants (s$^{-1}$). The steady-state concentration ([ROS]$_{ss}$) can be obtained by calculating Eqs. 1-3.

*2.7. Computational methods*

The Fukui function based on Density Function Theory (DFT) calculation was calculated to predict the active sites of possible attack sites on TC molecule[24], which is defined as follows Eq. 4:

$$f(r) = [\frac{\partial \mu}{\partial v(r)}]_N = [\frac{\partial \rho(r)}{\partial N}]_{v(r)} \quad (4)$$

where $\rho(r)$ is the electron density at a point $r$ in space, $N$ is the electron number in the current system, the constant term $v$ in the partial derivative is external potential. In the condensed version of Fukui function, atomic population number is utilized to represent the amount of electron density distribution around an atom[25]. The condensed Fukui function ($f$) is calculated using Eqs. 5-7:

Nucleophilic attack: $f_A^+ = q_N^A - q_{N+1}^A$ (5)

Electrophilic attack: $f_A^- = q_{N-1}^A - q_N^A$ (6)

Radical attack: $f_A^0 = (q_{N-1}^A - q_{N+1}^A)/2$ (7)

Similarly, condensed dual descriptor ($\Delta f$) can be written using Eq. 8:

$$\Delta f_A = f_A^+ - f_A^- = 2q_N^A - q_{N+1}^A - q_{N-1}^A \quad (8)$$

where $q^A$ is the atomic charge of atom A at the corresponding state. In this study, Hirshfeld charge was employed to investigate reactive sites. Toxicity Estimation Software Tool (TEST) was utilized to investigate the toxicity of intermediates in the process of TC degradation.

## 3. Results and discussion

*3.1 Catalyst characterization*

After synthesis, SEM and TEM were used to investigate the morphology of as-prepared $Co_3O_4$, $NiCo_2O_4$ and $NiCo_2O_4$@C. As shown in Fig. 2a, the original ZIF-67 exhibited a solid dodecahedron shape with a size of about 1 μm. After calcination, the size of $Co_3O_4$, $NiCo_2O_4$ and $NiCo_2O_4$@C decreased to varying extents (Figs. 2b-d), and TEM images (Figs. 2e-f) of $NiCo_2O_4$@C indicated the formation of a hollow nanocage structure. The smaller size and hollow structure increased the specific surface area of the $NiCo_2O_4$@C, exposing more reactive centers for TC degradation. Interestingly, the shrinkage behaviors differed among different catalysts. While $Co_3O_4$ and $NiCo_2O_4$@C inherited the dodecahedral shape of ZIF-67, $NiCo_2O_4$ transformed into a cube. It could be speculated that pre-carbonization helped maintain the dodecahedral framework, preventing excessive shrinkage and collapse during calcination. Another essential factor that could lead to structure collapse is the ramping rate. As shown in Fig. S1, when the calcination of $NiCo_2O_4$@C was performed at a heating rate of 2 °C min$^{-1}$ and 5 °C min$^{-1}$, more intense shrinkage resulted in severe deformation and collapse, potentially reducing effective reactive sites. These findings confirmed the successful construction of the unique hollow nanocage structure of $NiCo_2O_4$@C.

To preliminarily reveal the composition of $NiCo_2O_4$@C hollow nanocages, HRTEM (Fig. 2g) was used to identify the crystal phases. The lattice fringe spacing of 0.203 nm and 0.245 nm corresponded to (400) and (311) planes of $NiCo_2O_4$, respectively. SAED patterns (Fig. 2h) also showed lattice spacing of 0.245 nm, 0.203 nm and 0.144 nm, corresponding to the (311), (400) and (440) planes of $NiCo_2O_4$. Furthermore, elemental mapping in Fig. 2i suggested that Ni, Co, O and C were uniformly distributed on the nanocages, albeit the content of C was lower than the others. The results confirmed the intense combination of in-situ carbon and $NiCo_2O_4$.

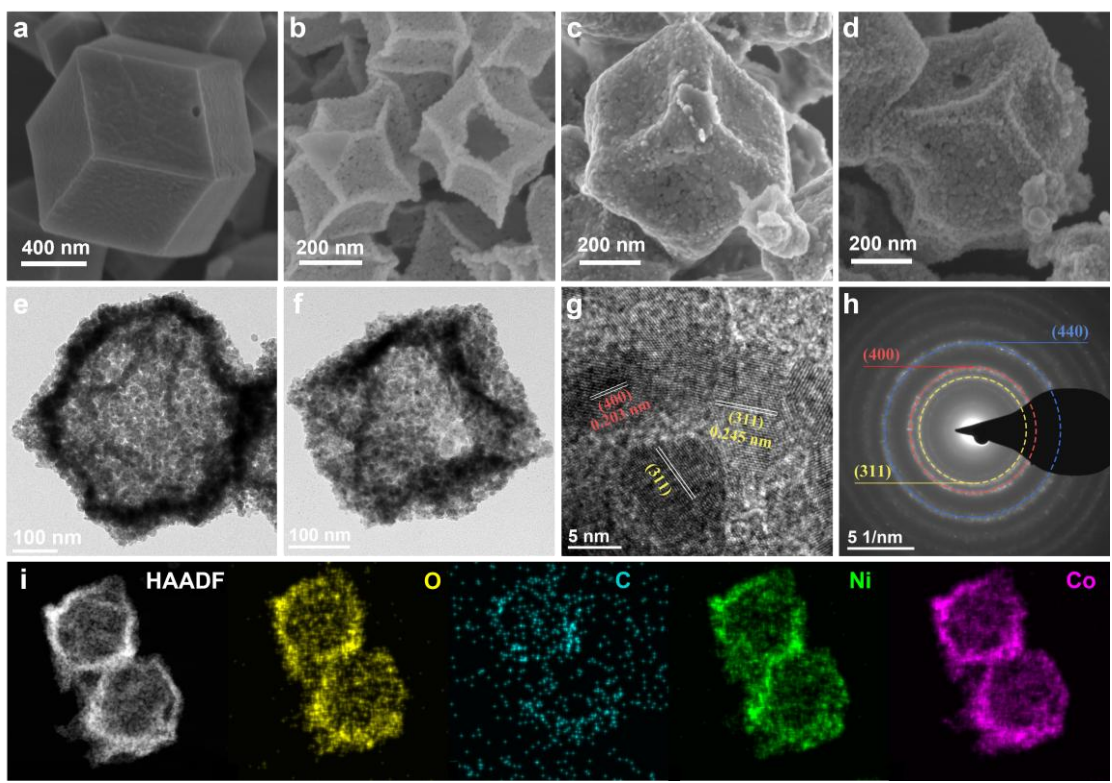

**Fig. 2**. SEM images of (a) ZIF-67, (b) Co₃O₄, (c) NiCo₂O₄ and (d) NiCo₂O₄@C. (e-f) TEM images, (g) HRTEM, (h) SAED, and (i) elemental mapping of NiCo₂O₄@C hollow nanocages.

XRD was employed to confirm the composition of different catalysts further. As shown in Fig. 3a, the diffraction peaks of Co₃O₄ could be indexed to cubic Co₃O₄ (PDF #74-2120), and the peaks of NiCo₂O₄ and NiCo₂O₄@C matched cubic NiCo₂O₄ (PDF #02-1074), which coincides with the results of HRTEM. Notably, the signal peaks of NiCo₂O₄@C were broader and weaker than those of Co₃O₄ and NiCo₂O₄, suggesting a smaller crystal size in NiCo₂O₄@C. Raman spectroscopy was further employed to verify the existence of carbon. In Fig. 3b, the peaks at 1346 cm⁻¹ and 1600 cm⁻¹ were ascribed to the D band and G band of carbon, respectively, indicating the existence of both sp³ and sp² hybridized carbon. Raman spectra of three catalysts (Fig. S2) all showed peaks at 190, 474, 516 and 681 cm⁻¹, corresponding to the vibration mode of $F^1_{2g}$, $E_g$, $F^2_{2g}$ and $A_{1g}$ in spinel oxides, respectively. Besides, FT-IR (Fig. S3) also confirmed the formation of spinel structure in all prepared catalysts. These results suggested the successful preparation of spinel oxides with in-situ generated carbon. Specific surface area was measured by the BET method (Fig. 3c and Table S3). The BET isotherms of all the catalysts were identified as type IV, indicative of a mesoporous structure. Moreover, the pore size distribution in Fig. 3d indicates that Ni etching and carbonization significantly reduced the mesopore diameter from approximately 22.2 nm to 10 nm. The BET specific surface area of $Co_3O_4$, $NiCo_2O_4$ and $NiCo_2O_4@C$ were 53.89, 75.77 and 62.9 $m^2\ g^{-1}$, respectively. Above all, $NiCo_2O_4@C$ hollow nanocages were successfully fabricated.

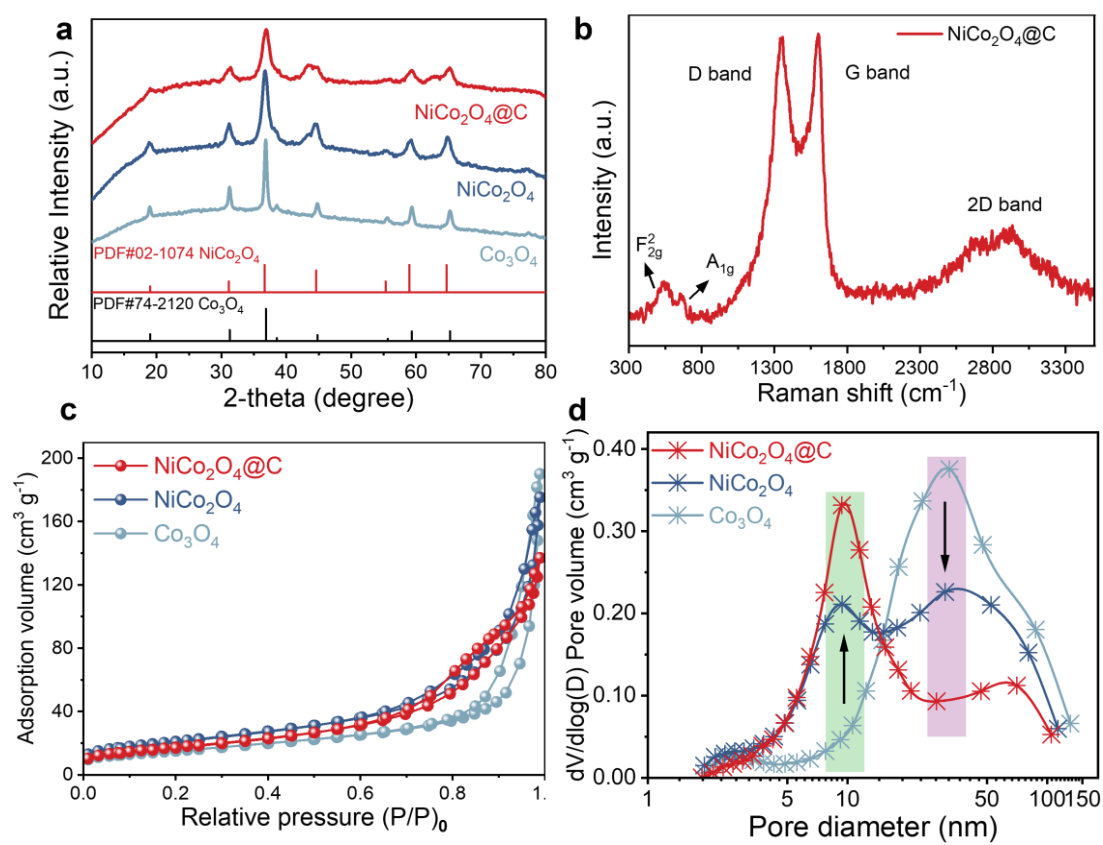

**Fig. 3.** (a) XRD patterns, (b) Raman spectra of $NiCo_2O_4@C$, (c) $N_2$ adsorption-desorption isotherms, and (d) pore diameter distribution of $Co_3O_4$, $NiCo_2O_4$ and $NiCo_2O_4@C$.

To unveil the inherent catalytic activity of NiCo$_2$O$_4$@C, the surface element valence state of the catalysts was investigated by XPS. Fig. 4a shows Co 2p spectra of Co$_3$O$_4$, NiCo$_2$O$_4$ and NiCo$_2$O$_4$@C. Peaks at 779.96 eV and 795.26 eV were attributed to Co$^{3+}$ 2p$_{3/2}$ and Co$^{3+}$ 2p$_{1/2}$, and peaks at 781.36 eV and 797.09 eV belonged to Co$^{2+}$ 2p$_{3/2}$ and Co$^{2+}$ 2p$_{1/2}$, respectively. The ratios of Co$^{2+}$/Co$^{3+}$ in Co$_3$O$_4$, NiCo$_2$O$_4$ and NiCo$_2$O$_4$@C were 0.87, 1.11 and 1.28, respectively. Namely, the content of Co$^{2+}$ exceeded Co$^{3+}$ in NiCo$_2$O$_4$ and NiCo$_2$O$_4$@C and was higher than that in Co$_3$O$_4$. Therefore, the increasing Co$^{2+}$ species in NiCo$_2$O$_4$ and NiCo$_2$O$_4$@C could accelerate PMS activation and TC degradation efficiency. The electron transfer between Ni$^{2+}$ and Co$^{3+}$ accounted for the increase in Co$^{2+}$ content. Since the redox potential of Ni$^{3+}$/Ni$^{2+}$ and Co$^{3+}$/Co$^{2+}$ were 3.7 V and 1.5 V [26,27], the reduced Co$^{3+}$ by Ni$^{2+}$ was thermodynamically available. Ni 2p spectra in Fig. 4b confirmed the apparent existence of Ni$^{3+}$ (856.16 eV). Overwhelmingly, it was reported that Ni$^{2+}$ tends to occupy the octahedral site in spinel oxide [28]. Thus, the edge-sharing interaction between Co$^{3+}$$_{Oct}$ and Ni$^{2+}$$_{Oct}$ was more reasonable (Fig. 4c).

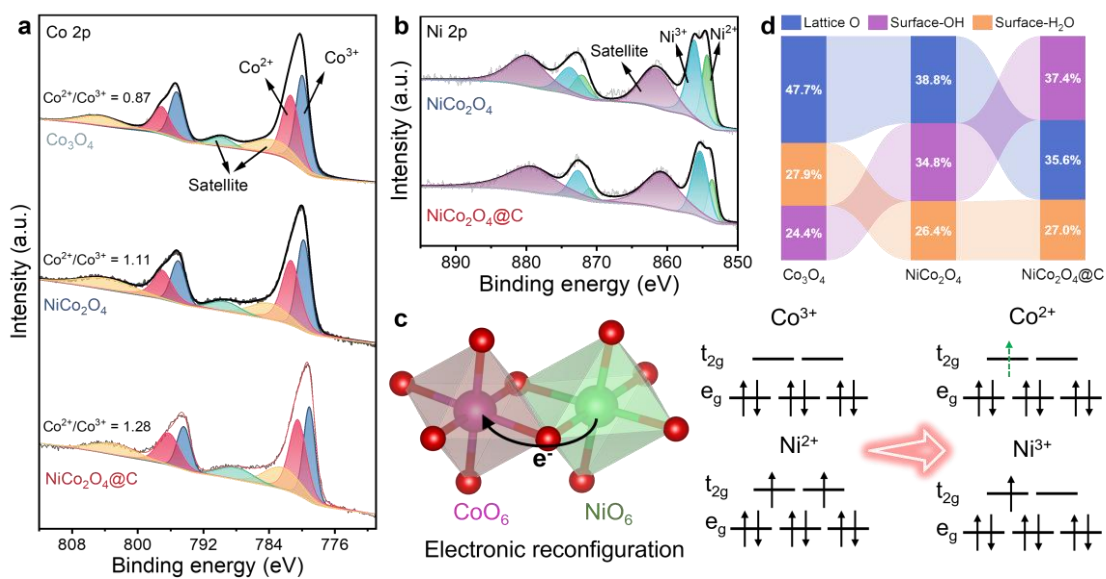

**Fig. 4.** (a) Co 2p, (b) Ni 2p XPS spectra, and (c) representative illustration of electron configuration between octahedral Co and Ni. (d) Oxygen species distribution of $Co_3O_4$, $NiCo_2O_4$ and $NiCo_2O_4$@C.

Calculated by O 1s spectra (Fig. S4), the distribution of lattice oxygen (Lattice O), adsorbed hydroxyl group (Surface-OH), and adsorbed $H_2O$ (Surface-$H_2O$) (Fig. 4d) clearly showed that the content of Surface-OH increased by more than 10% in $NiCo_2O_4$ and $NiCo_2O_4$@C compared to $Co_3O_4$, while that of lattice oxygen decreased by a similar amount. The increasing adsorbed hydroxyl groups on the surface decreased the activation barrier for water dissociation, and thus catalysts became more hydrophilic [29]. $NiCo_2O_4$ (18.6°) and $NiCo_2O_4$@C (11.2°) exhibited lower contact angles than $Co_3O_4$ (36.8°) in sessile drop analysis, indicating that the catalysts' hydrophilicity was enhanced (Fig. S5). The higher hydrophilicity benefited the adsorption of TC since it is hydrophilic [30], which consequently improved the efficiency of TC degradation.

*3.2 Catalytic performance*

TC adsorption experiments were firstly conducted to compare the adsorption capacity of different catalysts. As depicted in Fig. S6, $NiCo_2O_4$@C hollow nanocages could adsorb about 20% of TC within 60 min, 4 and 2 times higher than $Co_3O_4$ and $NiCo_2O_4$, respectively. These results confirmed the enhanced hydrophilicity had improved the adsorption of TC. On the other hand, despite $NiCo_2O_4$ having a larger surface area than $NiCo_2O_4$@C, the presence of carbon, which is electron-rich, enables higher adsorption capacity due to the strong electrostatic adsorption of TC.

The degradation performance of $NiCo_2O_4$@C was investigated using 10 mg $L^{-1}$ TC as the target pollutant. As shown in Fig. 5a, pure PMS degraded 18% of TC due to self-activation. In the $Co_3O_4$, $NiCo_2O_4$, and $NiCo_2O_4$@C systems, TC degradation efficiencies attained 85.2%, 93.0% and 98.1% within 20 min, respectively. The corresponding kinetic rate constants were calculated to be 0.0905, 0.1249 and 0.1829 min$^{-1}$ (Fig. 5b). Although both NiCo$_2$O$_4$ and NiCo$_2$O$_4$@C systems removed more than 90% of TC, the kinetic constant of the latter was 1.46 times faster. This can be attributed to the extraordinary TC adsorption capabilities of NiCo$_2$O$_4$@C, which facilitated the removal of pollutants by shortening the distance between TC and ROS. Metal leaching in the NiCo$_2$O$_4$ and NiCo$_2$O$_4$@C systems was detected using ICP-OES to evaluate catalyst stability. As shown in Fig. 5c, Co and Ni leaching were significantly inhibited in the NiCo$_2$O$_4$@C systems. In particular, the leaching level of Ni decreased from 20.09 mg L$^{-1}$ to 4.18 mg L$^{-1}$. These results revealed that carbon in NiCo$_2$O$_4$@C could stabilize the spinel oxides. Additionally, XRD and XPS spectra of NiCo$_2$O$_4$@C before and after reactions (Figs. S7-S8) showed minor changes in crystal phases or in chemical states of Ni and Co, implying excellent stability. An equal amount of leaching Ni and Co ions in the NiCo$_2$O$_4$@C system were added directly to activate PMS for TC degradation. As shown in Fig. 5a, negligible catalytic performance could be observed in the Ni$^{2+}$ and Co$^{2+}$ system compared to the PMS alone system, excluding the impact of homogenous activation of PMS by Co and Ni ions. The capability of NiCo$_2$O$_4$@C was further evaluated for the degradation of other tetracycline antibiotics, including chlortetracycline (CTC) and oxytetracycline (OTC). Fig. 5d exhibited 74.8% and 83.5% removal efficiencies for CTC and OTC, respectively, demonstrating satisfactory capability for degrading other tetracycline antibiotics. Fig. 5e and Table S4 compared the TC degradation performance of the optimal NiCo$_2$O$_4$@C/PMS system in this study with previously reported catalyst/PMS systems [31–38] based on kinetic rate, catalyst dosage, and PMS/TC ratio. The results above demonstrated that the NiCo$_2$O$_4$@C hollow nanocages possess excellent catalytic activity and stability.

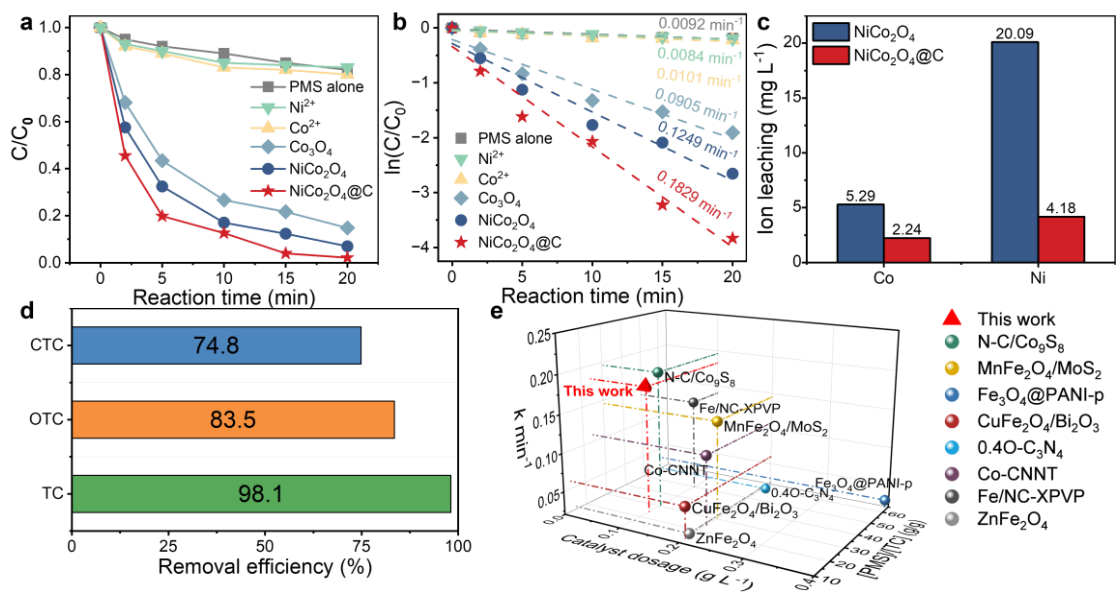

**Fig. 5**. (a) TC removal in different reaction systems and (b) corresponding kinetic constant. (c) Co and Ni leaching concentration in the $NiCo_2O_4$/PMS and $NiCo_2O_4$@C/PMS systems. (d) TC, OTC, and CTC removal efficiencies in the $NiCo_2O_4$@C/PMS systems. (e) A comparison of TC degradation performance of the $NiCo_2O_4$@C/PMS system and other catalyst/PMS systems [31–38]. Experimental conditions: [catalyst] = 0.1 g L$^{-1}$, [TC] = 10 mg L$^{-1}$, [PMS] = 0.8 mM, initial pH = 7.0, T = 298 K.

Commonly existing anions and cations in the water matrix, as well as pH, can impact the degradation of organic pollutants. Therefore, we conducted various environmental impact factor tests to evaluate the anti-interference capability of the $NiCo_2O_4$@C/PMS system. In Fig. S9a, Na$^+$, K$^+$, Mg$^{2+}$ and Ca$^{2+}$ harmed TC degradation, while the inhibition effect was more evident with monovalent cations. Fig. S9b displayed the influence of anions (Cl$^-$, NO$_3^-$, SO$_4^{2-}$ and CO$_3^{2-}$). While Cl$^-$, NO$_3^-$, and SO$_4^{2}$ limited the rate constant of TC degradation, CO$_3^{2-}$ enhanced the performance of TC degradation, resulting in the removal of more than 90% of TC within 5 min. It has been reported that dissolved carbonates can enhance catalytic activity by producing ·CO$_3^-$ radicals and serving as electron mediators for the redox cycle of metals [39]. Moreover, the effect of pH on TC degradation was also studied. Fig. S9c suggested that the optimal pH was 7, while about 80% of TC could still be removed under acidic or basic conditions. These results confirmed the exceptional ecological resistance of the NiCo$_2$O$_4$@C/PMS system.

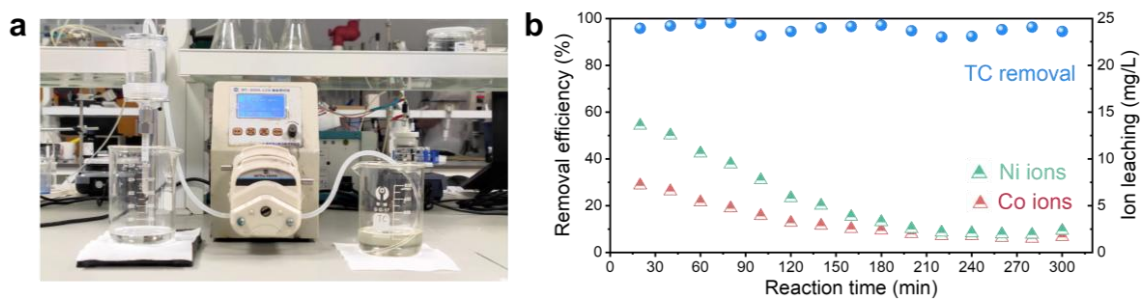

**Fig. 6**. (a) Schematic illustration of the simulated continuous flow system. (b) TC degradation performance and metal leaching detection in the continuous flow system. Experimental conditions: Catalyst loading = 50 mg, [TC] = 10 mg L$^{-1}$, [PMS] = 0.8 mM, flow rate = 5 mL min$^{-1}$, initial pH = 7.0, T = 298 K.

To evaluate the reusability and potential application in continuous operating conditions, a constant flow system was built to monitor the catalytic performance of NiCo$_2$O$_4$@C hollow nanocages (Fig. 6a). 50 mg catalysts were loaded on the PTFE membrane by vacuum filtration, and the TC degradation process was run for 300 min. As shown in Fig. 6b, TC removal efficiency sustained more than 90% during continuous operation. Moreover, the leaching of Ni and Co decreased while maintaining excellent performance, suggesting good reusability and less metal leaching pollution to the environment. In general, the NiCo$_2$O$_4$@C/PMS system exhibited superior catalytic performance in TC degradation.

*3.3 Mechanisms*

*3.3.1 Determination of ROS*

Scavenging experiments were first conducted to quantitatively determine the contribution of different ROS for TC removal in the NiCo$_2$O$_4$@C/PMS system. The commonly seen ROS in PMS activation included ·OH, ·SO$_4^-$, ·O$_2^-$ and $^1$O$_2$. Therefore, methanol (MeOH), tert-butanol (TBA), furfuryl alcohol (FFA) and p-benzoquinone (p-BQ) were used to scavenge various ROS and identify the dominant species. As shown in Fig. 7a, p-BQ quenching test revealed the negligible effect of ·O$_2^-$, while the FFA quenching test suggested the vital participation of $^1$O$_2$. MeOH and TBA only showed slight inhibition of TC degradation, indicating the minor role of ·OH and ·SO$_4^-$ in the reaction system. EPR was utilized to detect the existence of ROS directly with DMPO and TEMP as spin-trapping reagents (Figs. 7b-d). Septet signals of DMPOX were seen in the NiCo$_2$O$_4$@C/PMS system, which may be due to the direct oxidation of DMPO by PMS [40]. No signals of DMPO-·O$_2^-$ were detected, and robust triple peaks of TEMP-$^1$O$_2$ were detected. These results suggested that $^1$O$_2$ was the main ROS.

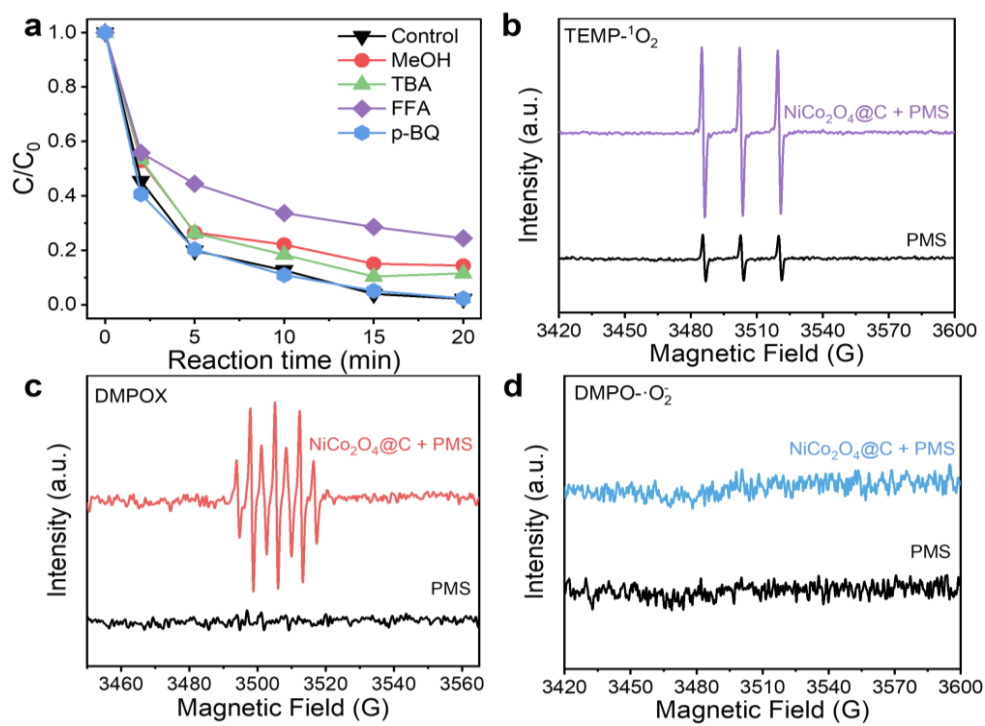

**Fig. 7**. (a) Quenching experiments of TC degradation and (b) corresponding kinetic rates in the NiCo$_2$O$_4$@C/PMS system. (c) Calculated contributions of different ROS in the NiCo$_2$O$_4$@C/PMS system. EPR spectra captured by (d-e) DMPO and (f) TEMP. Experimental conditions: [MeOH] = [TBA] = 100 mM, [FFA] = [p-BQ] = 10 mM, [catalyst] = 0.1 g L$^{-1}$, [TC] = 10 mg L$^{-1}$, [PMS] = 0.8 mM, initial pH = 7.0, T = 298 K.

Chemical probes, including benzoic acid (BA), nitrobenzene (NB) and FFA, were employed to further quantify the steady-state concentration of different ROS (Fig. 8). Calculation results showed that the steady-state concentration of $^1$O$_2$ (2.13 × 10$^{-11}$ M) was significantly larger than that of ·OH (9.40 × 10$^{-14}$ M) and ·SO$_4^-$ (1.22 × 10$^{-14}$ M), which was consistent with the quenching experiments. However, the addition of FFA only inhibited 22% of TC removal efficiency, indicating the essential participation of direct electron transfer reaction pathway.

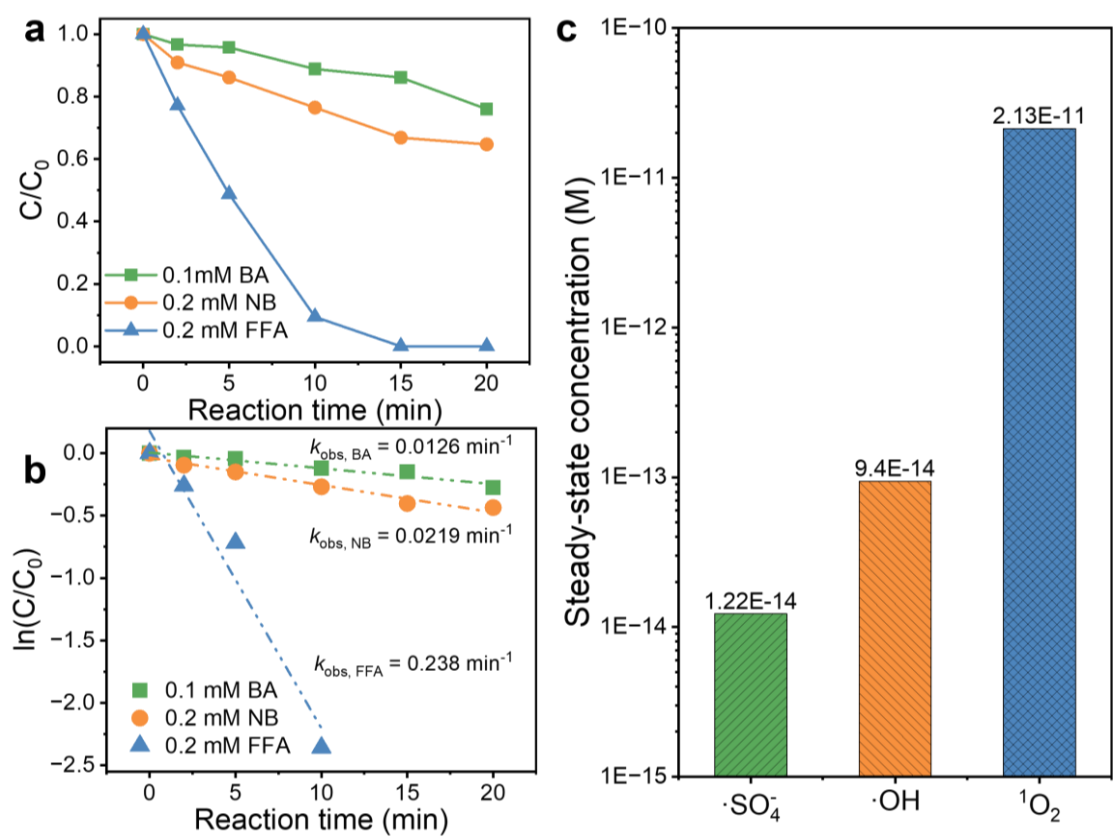

**Fig. 8.** (a) The degradation of chemical probes in the NiCo$_2$O$_4$@C/PMS system and (b) corresponding pseudo-first-order rate constant. (c) Steady-state concentration of ·OH, ·SO$_4^-$, and $^1$O$_2$.

*3.3.2. Electron-shuttle effect of in-situ carbon*

The electrical conductivity of catalysts was measured via the four-point probe method. As shown in Fig. 9a, the conductivity of $Co_3O_4$, $NiCo_2O_4$ and $NiCo_2O_4$@C were determined to be 16.79, 30.99, and 201.82 S m$^{-1}$. Compared with $Co_3O_4$, the substitution of Ni for Co in $NiCo_2O_4$ facilitated the electron transfer and exhibited a synergistic effect between Ni and Co for PMS activation. Moreover, a surge of conductivity appeared in the $NiCo_2O_4$@C due to the existence of in-situ carbon. LSV curves further confirmed the apparent increase of the ETR in $NiCo_2O_4$@C. In-situ carbon with abundant free electrons serves as the "electron shuttles", mediating the electron transfer between $NiCo_2O_4$ and PMS as well as the redox cycle between Co and Ni, which enhanced the activity and stability of $NiCo_2O_4$@C hollow nanocages. Moreover, chronoamperometry measurements were conducted to confirm the direct electron transfer reaction pathway (Fig. 9b). Compared to the other two catalysts, the injection of PMS and TC caused distinct current jumps for the $NiCo_2O_4$@C, demonstrating efficient electron transfer between $NiCo_2O_4$@C and PMS. The result also confirmed the essential contribution of direct electron transfer pathway in the $NiCo_2O_4$@C/PMS system (Fig. 9c) [41,42]. The above discussion collectively demonstrated that the notable activity of the $NiCo_2O_4$@C/PMS system was due to the enhanced ETR, which resulted from the combined effects of the nanocage confinement, Ni-Co synergy, and the electron-shuttle effect.

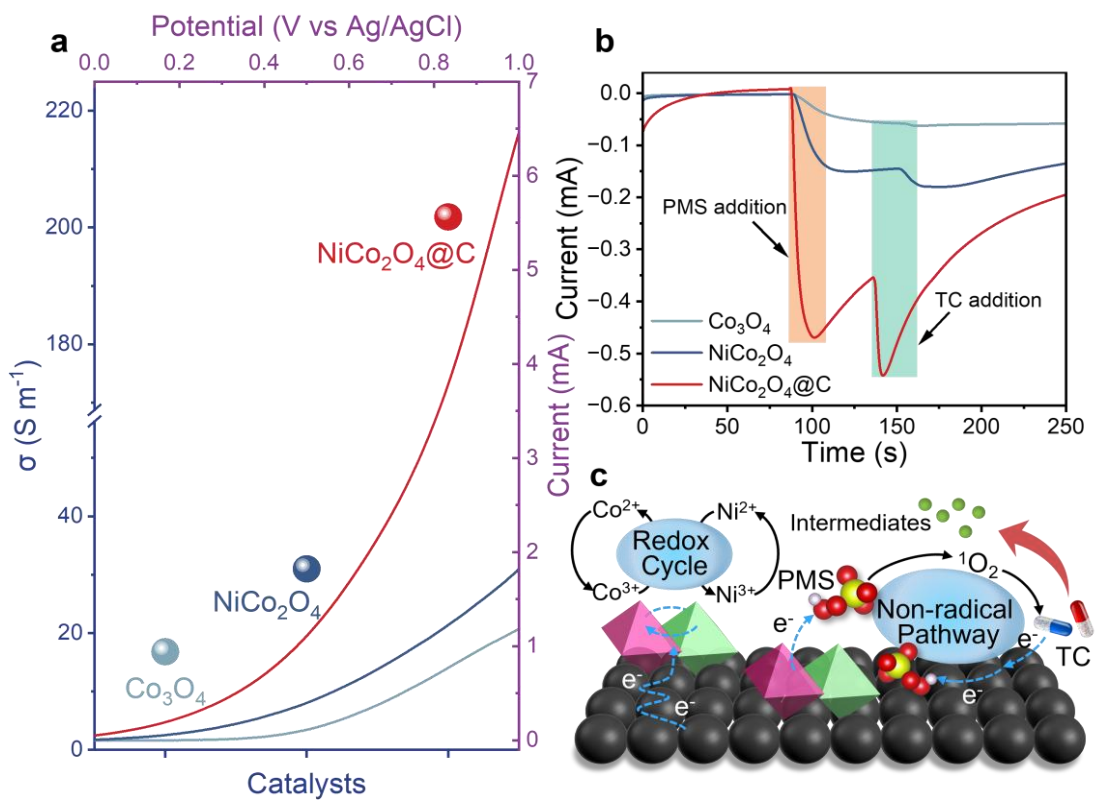

**Fig. 9**. (a) Conductivity and LSV curves, and (b) i-t curves of $Co_3O_4$, $NiCo_2O_4$ and $NiCo_2O_4$@C under 0.4 V vs Ag/AgCl with 1 M $Na_2SO_4$ as the electrolyte. (c) Proposed degradation mechanism.

*3.4 Degradation intermediates and toxicity*

To identify possible degradation intermediates, DFT calculations were utilized to predicted attack sites on TC (Table S5). Fukui index predicted the radical and non-radical attack sites on TC. Fig. 10a suggested that the most positive part of $f^0$ function was mainly on C3, O22 and O23, related to the major carbonyl group, which suggests radical attack tended to happen on these sites. The most positive part of $f^-$ function (Fig. 10b) was mainly on C16, C20, O22, O26 and O27, related to the hydroxyl group and amino group on TC, indicating electrophilic attack may happen on these sites. Possible attack sites on TC were demonstrated in Fig. 10c. Noticeably, the number of $f^-$ function values higher than 0.04 was more than that of $f^0$ and $f^+$ (Fig. 10d), suggesting that electrophilic attack was more likely to happen on TC.

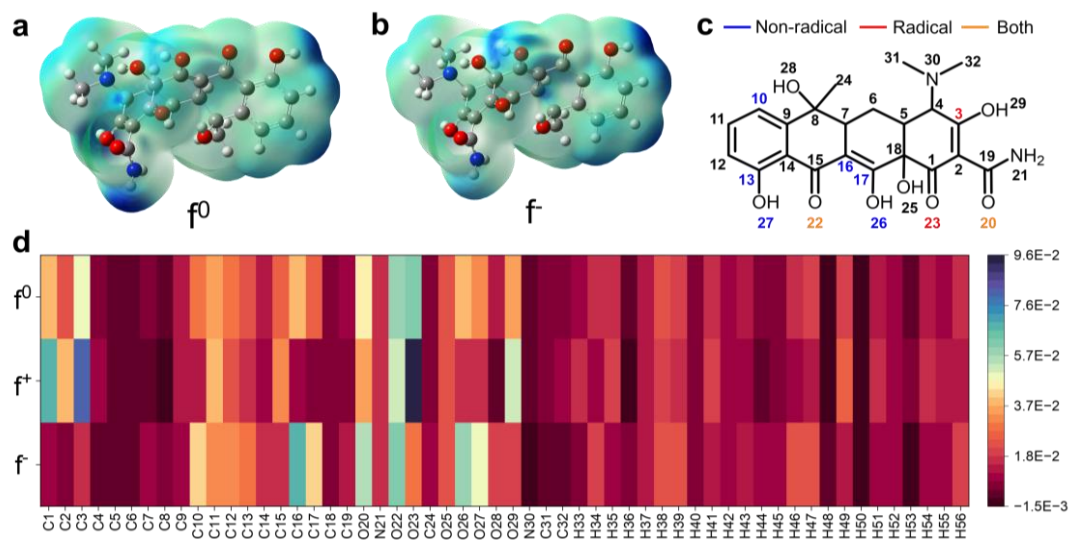

**Fig. 10.** Visualization of (a) $f^0$ and (b) $f^-$ Fukui functions of TC. (c) Attack sites on TC molecule. (d) $f^0$, $f^+$ and $f^-$ Fukui function value distribution.

TC and its degradation intermediates were analyzed using LC-MS (Fig. S10), and the possible degradation pathway was proposed in Fig. 11a. With the dehydration and hydroxylation processes, TC (m/z = 445) could be transformed into P2 (m/z = 427) and P5 (m/z = 459). Through serial ring opening procedures, P2 was further degraded into P3 (m/z = 346) and P4 (m/z = 247). Due to removing the N-methyl substituent, P5 transformed into P6 (m/z = 374), which was further converted to P7 (m/z = 337) through dehydration and the ring opening reactions. Multiple hydrogenation and N-dealkylation reaction converted P5 to P8 (m/z = 365) and P10 (m/z = 362). Continuous ring opening reactions formed P9 (m/z = 337) and P11 (m/z = 274). Toxicity of TC and intermediates was predicted with TEST. As shown in Figs. 11b-c and Table S6, $LC_{50}$ (Lethal Concentration, 50%) of TC to fathead minnow and daphnia magna were 0.90 and 8.73 mg $L^{-1}$, respectively, identified as the toxicant. After the degradation process, although the toxicity of P2, P3, and P11 was higher than that of the original TC molecules, $LC_{50}$ of most intermediates increased significantly, suggesting the general abatement of toxicity. These results confirm that the degradation process could reduce biological toxicity of antibiotics.

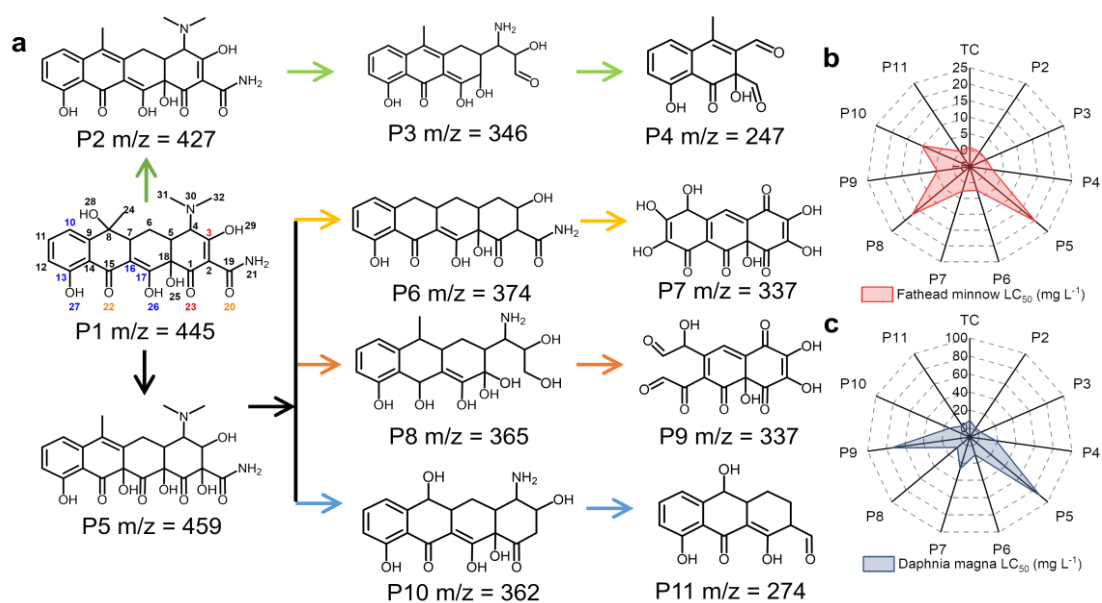

**Fig. 11.** (a) Possible degradation pathways. Acute toxicity LC$_{50}$ of (b) fathead minnow and (c) daphnia magna.

**4. Conclusions**

In this work, we have enhanced the electron transfer rate of the Fenton-like system via preparing ZIF-derived NiCo$_2$O$_4$@C hollow nanocages with in-situ generated carbon. The optimal NiCo$_2$O$_4$@C/PMS system exhibited a notable removal efficiency of 98.1% for 10 mg L$^{-1}$ TC within 20 min. Detailed characterization and experiments have demonstrated that the hollow nanocage structure promoted TC adsorption through confinement effect. The synergy between Ni and Co increased the ratio of Co$^{2+}$, resulting in more efficient PMS activation. In-situ carbon with abundant free electrons largely increased the electrical conductivity of NiCo$_2$O$_4$@C to 201.82 S m$^{-1}$, serving as electron shuttles to facilitate PMS activation and stabilize NiCo$_2$O$_4$. Non-radical reaction pathways were found to play a dominant role in the NiCo$_2$O$_4$@C/PMS system.

DFT calculations and toxicity evaluation indicated that the degradation intermediates exhibited reduced toxicity, effectively mitigating the environmental impact caused by antibiotics. Overall, the rationally designed hollow spinel oxides provide a distinctive approach to enhancing the efficiency of Fenton-like systems for antibiotic removal.

**Acknowledgments**

This work was supported by the National Key R&D Program of China (2023YFC3905804), the National Natural Science Foundation of China (22078374, 22378434), the Scientific and Technological Planning Project of Guangzhou (202206010145).

**References**

[1] S.C. Roberts, T.R. Zembower, Global increases in antibiotic consumption: A concerning trend for WHO targets, Lancet Infect. Dis. 21 (2021) 10-11. https://doi.org/10.1016/S1473-3099(20)30456-4.

[2] D.G.J. Larsson, C.-F. Flach, Antibiotic resistance in the environment, Nat. Rev. Microbiol. 20 (2022) 257–269. https://doi.org/10.1038/s41579-021-00649-x.

[3] R. Huang, P. Ding, D. Huang, F. Yang, Antibiotic pollution threatens public health in China, Lancet 385 (2015) 773–774. https://doi.org/10.1016/S0140-6736(15)60437-8.

[4] B.C. Hodges, E.L. Cates, J.-H. Kim, Challenges and prospects of advanced oxidation water treatment processes using catalytic nanomaterials, Nat. Nanotechnol. 13 (2018) 642–650. https://doi.org/10.1038/s41565-018-0216-x.

[5]  T. Liu, S. Xiao, N. Li, J. Chen, X. Zhou, Y. Qian, C.-H. Huang, Y. Zhang, Water decontamination via nonradical process by nanoconfined Fenton-like catalysts, Nat. Commun. 14 (2023) 2881. https://doi.org/10.1038/s41467-023-38677-1.

[6]  Q. Zhou, C. Song, P. Wang, Z. Zhao, Y. Li, S. Zhan, Generating dual-active species by triple-atom sites through peroxymonosulfate activation for treating micropollutants in complex water, Proc. Natl. Acad. Sci. U.S.A. 120 (2023) e2300085120. https://doi.org/10.1073/pnas.2300085120.

[7]  B.C. Hodges, E.L. Cates, J.-H. Kim, Challenges and prospects of advanced oxidation water treatment processes using catalytic nanomaterials, Nat. Nanotechnol. 13 (2018) 642–650. https://doi.org/10.1038/s41565-018-0216-x.

[8]  T. Liu, S. Xiao, N. Li, J. Chen, X. Zhou, Y. Qian, C.-H. Huang, Y. Zhang, Water decontamination via nonradical process by nanoconfined Fenton-like catalysts, Nat. Commun. 14 (2023) 2881. https://doi.org/10.1038/s41467-023-38677-1.

[9]  J. Lim, Y. Yang, M.R. Hoffmann, Activation of peroxymonosulfate by oxygen vacancies-enriched cobalt-doped black $TiO_2$ nanotubes for the removal of organic pollutants, Environ. Sci. Technol. 53 (2019) 6972–6980. https://doi.org/10.1021/acs.est.9b01449.

[10] Z. Yang, Y. Huang, X. Li, Z. Jiang, Y. Chen, S. Yang, H.F. Garces, Y. Sun, K. Yan, Highly dispersed CoFe2O4 spinel on biomass-derived 3D porous carbon framework for much enhanced Fenton-like reactions, Sep. Purif. Technol. 298 (2022) 121535. https://doi.org/10.1016/j.seppur.2022.121535.

[11] Z.-Y. Guo, Y. Si, W.-Q. Xia, F. Wang, H.-Q. Liu, C. Yang, W.-J. Zhang, W.-W. Li, Electron delocalization triggers nonradical Fenton-like catalysis over spinel oxides, Proc. Natl. Acad. Sci. U.S.A. 119 (2022) e2201607119. https://doi.org/10.1073/pnas.2201607119.

[12] J. Li, D. Chu, H. Dong, D.R. Baker, R. Jiang, Boosted oxygen evolution reactivity by igniting double exchange interaction in spinel oxides, J. Am. Chem. Soc. 142 (2020) 50–54. https://doi.org/10.1021/jacs.9b10882.

[13] S. Liu, B. Zhang, Y. Cao, H. Wang, Y. Zhang, S. Zhang, Y. Li, H. Gong, S. Liu, Z. Yang, J. Sun, Understanding the effect of nickel doping in cobalt spinel oxides on regulating spin state to promote the performance of the oxygen reduction reaction and zinc–air batteries, ACS Energy Lett. (2022) 159–168. https://doi.org/10.1021/acsenergylett.2c02457.

[14] J. Zhao, Y. He, J. Wang, J. Zhang, L. Qiu, Y. Chen, C. Zhong, X. Han, Y. Deng, W. Hu, Regulating metal active sites of atomically-thin nickel-doped spinel cobalt oxide toward enhanced oxygen electrocatalysis, Chem. Eng. J. 435 (2022) 134261. https://doi.org/10.1016/j.cej.2021.134261.

[15] Y.-L. He, C.-S. He, L.-D. Lai, P. Zhou, H. Zhang, L.-L. Li, Z.-K. Xiong, Y. Mu, Z.-C. Pan, G. Yao, B. Lai, Activating peroxymonosulfate by N and O co-doped porous carbon for efficient BPA degradation: A re-visit to the removal mechanism and the effects of surface unpaired electrons, Appl. Catal. B: Environ. 314 (2022) 121390. https://doi.org/10.1016/j.apcatb.2022.121390.

[16] Z. Wang, E. Almatrafi, H. Wang, H. Qin, W. Wang, L. Du, S. Chen, G. Zeng, P. Xu, Cobalt single atoms anchored on oxygen‐doped tubular carbon nitride for efficient peroxymonosulfate activation: simultaneous coordination structure and morphology modulation, Angew. Chem. Int. Ed. 61 (2022). https://doi.org/10.1002/anie.202202338.

[17] J. Zhang, H. Liu, W. Gao, D. Cheng, F. Tan, W. Wang, X. Wang, X. Qiao, P.K. Wong, Y. Yao, In situ zinc cyanamide coordination induced highly N-rich

graphene for efficient peroxymonosulfate activation, J. Mater. Chem. A 10 (2022) 12016–12025. https://doi.org/10.1039/D2TA00506A.

[18] L. Ji, F. Liu, Z. Xu, S. Zheng, D. Zhu, Adsorption of pharmaceutical antibiotics on template-synthesized ordered micro- and mesoporous carbons, Environ. Sci. Technol. 44 (2010) 3116–3122. https://doi.org/10.1021/es903716s.

[19] A. Jayakumar, R.P. Antony, R. Wang, J.-M. Lee, MOF-derived hollow cage $Ni_xCo_{3-x}O_4$ and their synergy with graphene for outstanding supercapacitors, Small 13 (2017) 1603102. https://doi.org/10.1002/smll.201603102.

[20] L. Qin, Z. Xu, Y. Zheng, C. Li, J. Mao, G. Zhang, Confined transformation of organometal-encapsulated MOFs into spinel $CoFe_2O_4$/C nanocubes for low-temperature catalytic oxidation, Adv. Funct. Mater. 30 (2020) 1910257. https://doi.org/10.1002/adfm.201910257.

[21] X. Guo, H. Zhang, Y. Yao, C. Xiao, X. Yan, K. Chen, J. Qi, Y. Zhou, Z. Zhu, X. Sun, J. Li, Derivatives of two-dimensional MXene-MOFs heterostructure for boosting peroxymonosulfate activation: Enhanced performance and synergistic mechanism, Appl. Catal. B: Environ. 323 (2023) 122136. https://doi.org/10.1016/j.apcatb.2022.122136.

[22] D. Saliba, M. Ammar, M. Rammal, M. Al-Ghoul, M. Hmadeh, Crystal growth of ZIF-8, ZIF-67, and their mixed-metal derivatives, J. Am. Chem. Soc. 140 (2018) 1812–1823. https://doi.org/10.1021/jacs.7b11589.

[23] Q.-Y. Wu, Z.-W. Yang, Z.-W. Wang, W.-L. Wang, Oxygen doping of cobalt-single-atom coordination enhances peroxymonosulfate activation and high-valent cobalt–oxo species formation, Proc. Natl. Acad. Sci. U.S.A. 120 (2023) e2219923120. https://doi.org/10.1073/pnas.2219923120.

[24] M.L. Cerón, T. Gomez, M. Calatayud, C. Cárdenas, Computing the Fukui Function in solid-state chemistry: application to alkaline earth oxides bulk and surfaces, J. Phys. Chem. A 124 (2020) 2826–2833. https://doi.org/10.1021/acs.jpca.0c00950.

[25] Z.-J. Xiao, X.-C. Feng, H.-T. Shi, B.-Q. Zhou, W.-Q. Wang, N.-Q. Ren, Why the cooperation of radical and non-radical pathways in PMS system leads to a higher efficiency than a single pathway in tetracycline degradation, J. Hazard. Mater. 424 (2022) 127247. https://doi.org/10.1016/j.jhazmat.2021.127247.

[26] J.-X. Huang, B. Li, B. Liu, B.-J. Liu, J.-B. Zhao, B. Ren, Structural evolution of NM (Ni and Mn) lithium-rich layered material revealed by in-situ electrochemical Raman spectroscopic study, J. Power Sources 310 (2016) 85–90. https://doi.org/10.1016/j.jpowsour.2016.01.065.

[27] A.L. Rotinjan, L.M. Borisowa, R.W. Boldin, Das redox-standardpotential von $Co^{3+}/Co^{2+}$, Electrochim. Acta 19 (1974) 43–46. https://doi.org/10.1016/0013-4686(74)80005-8.

[28] Z. Wang, P. Wu, X. Zou, S. Wang, L. Du, T. Ouyang, Z.-Q. Liu, Optimizing the oxygen-catalytic performance of Zn–Mn–Co spinel by regulating the bond competition at octahedral sites, Adv. Funct. Mater. 33 (2023) 2214275. https://doi.org/10.1002/adfm.202214275.

[29] S. Yamamoto, H. Bluhm, K. Andersson, G. Ketteler, H. Ogasawara, M. Salmeron, A. Nilsson, In situ x-ray photoelectron spectroscopy studies of water on metals and oxides at ambient conditions, J. Phys.: Condens. Matter 20 (2008) 184025. https://doi.org/10.1088/0953-8984/20/18/184025.

[30] S. Kang, H. Zhu, L. Wang, X. Zhao, X. Huang, H. Yang, M. Dou, D. Li, J. Dou, Adjusting hydrophilicity of g-$C_3N_4$ based heterojunction photocatalyst through

sulfur-impregnation to enhancing degradation effect of tetracycline, Mater. Sci. Semicond. Process. 158 (2023) 107351. https://doi.org/10.1016/j.mssp.2023.107351.

[31] G. Zhang, J. Gao, J. Wang, H. Lin, J. Xu, L. Wang, ZIF-67/melamine derived hollow N-doped carbon/$Co_9S_8$ polyhedron to activate peroxymonosulfate for degradation of tetracycline, J. Environ. Chem. Eng. 11 (2023) 109355. https://doi.org/10.1016/j.jece.2023.109355.

[32] P. Xu, S. Xie, X. Liu, L. Wang, R. Wu, B. Hou, Efficient removal of tetracycline using magnetic $MnFe_2O_4$/$MoS_2$ nanocomposite activated peroxymonosulfate: Mechanistic insights and performance evaluation, Chem. Eng. J. 480 (2024) 148233. https://doi.org/10.1016/j.cej.2023.148233.

[33] Y. Wang, K. Li, M. Shang, Y. Zhang, Y. Zhang, B. Li, Y. Kan, X. Cao, J. Zhang, A novel partially carbonized $Fe_3O_4$@PANI-p catalyst for tetracycline degradation via peroxymonosulfate activation, Chem. Eng. J. 451 (2023) 138655. https://doi.org/10.1016/j.cej.2022.138655.

[34] Y. Li, C. Zhu, L. Chen, L. Liu, J. Zhang, N. Yang, Y. Li, Self-driven degradation of TC-HCl by $CuFe_2O_4$/$Bi_2O_3$ activated peroxymonosulfate, Chem. Eng. J. 473 (2023) 145282. https://doi.org/10.1016/j.cej.2023.145282.

[35] L. Wang, R. Li, Y. Zhang, Y. Gao, X. Xiao, Z. Zhang, T. Chen, Y. Zhao, Tetracycline degradation mechanism of peroxymonosulfate activated by oxygen-doped carbon nitride, RSC Adv. 13 (2023) 6368–6377. https://doi.org/10.1039/D3RA00345K.

[36] B. Xu, X. Zhang, Y. Zhang, S. Wang, P. Yu, Y. Sun, X. Li, Y. Xu, Enhanced electron transfer-based nonradical activation of peroxymonosulfate by $CoN_x$ sites

anchored on carbon nitride nanotubes for the removal of organic pollutants, Chem. Eng. J. 466 (2023) 143155. https://doi.org/10.1016/j.cej.2023.143155.

[37] H. Li, N. Wang, H. Li, Z. Ren, W. Ma, J. Li, Y. Du, Q. Xu, Polyvinylpyrrolidone-induced size-dependent catalytic behavior of Fe sites on N-doped carbon substrate and mechanism conversion in Fenton-like oxidation reaction, Appl. Catal. B: Environ. 341 (2024) 123323. https://doi.org/10.1016/j.apcatb.2023.123323.

[38] Y. Wu, X. Wang, T. She, T. Li, Y. Wang, Z. Xu, X. Jin, H. Song, S. Yang, S. Li, S. Yan, H. He, L. Zhang, Z. Zou, Iron 3D-orbital configuration dependent electron transfer for efficient Fenton-like catalysis, Small 20 (2024) 2306464. https://doi.org/10.1002/smll.202306464.

[39] X. Wang, Y. Zhou, N. Wang, J. Zhang, L. Zhu, Carbonate-induced enhancement of phenols degradation in CuS/peroxymonosulfate system: A clear correlation between this enhancement and electronic effects of phenols substituents, J. Environ. Sci. 129 (2023) 139–151. https://doi.org/10.1016/j.jes.2022.09.018.

[40] Y. Han, C. Zhao, W. Zhang, Z. Liu, Z. Li, F. Han, M. Zhang, F. Xu, W. Zhou, Cu-doped CoOOH activates peroxymonosulfate to generate high-valent cobalt-oxo species to degrade organic pollutants in saline environments, Appl. Catal. B: Environ. 340 (2024) 123224. https://doi.org/10.1016/j.apcatb.2023.123224.

[41] K. Zhu, W. Qin, Y. Gan, Y. Huang, Z. Jiang, Y. Chen, X. Li, K. Yan, Acceleration of $Fe^{3+}/Fe^{2+}$ cycle in garland-like MIL-101(Fe)/MoS$_2$ nanosheets to promote peroxymonosulfate activation for sulfamethoxazole degradation, Chem. Eng. J. 470 (2023) 144190. https://doi.org/10.1016/j.cej.2023.144190.

[42] X. Zhou, M.-K. Ke, G.-X. Huang, C. Chen, W. Chen, K. Liang, Y. Qu, J. Yang, Y. Wang, F. Li, H.-Q. Yu, Y. Wu, Identification of Fenton-like active Cu sites by

heteroatom modulation of electronic density, Proc. Natl. Acad. Sci. U.S.A. 119 (2022) e2119492119. https://doi.org/10.1073/pnas.2119492119.